# Behavioural Sciences and the Regulation of Privacy on the Internet

**Frederik Zuiderveen Borgesius**

Frederikzb[at]cs.ru.nl

## Table of Contents

## 1 Introduction

This chapter examines the policy implications of behavioural sciences insights for the regulation of privacy on the Internet, by focusing in particular on behavioural targeting.[1] This marketing technique involves tracking people's online behaviour to use the collected information to show people individually targeted advertisements.

This chapter distinguishes two ways in which policymakers can defend privacy in the area of behavioural targeting. Policymakers can aim to *empower*, or to *protect* the individual.[2] First, policymakers can aim to empower the individual, for instance by enabling her to make choices in her own interests. In other words, policymakers can try to put people in control over their personal information. The European legal regime regarding privacy in the area of behavioural targeting largely aims at individual empowerment. For example, the e-Privacy Directive requires firms to obtain the individual's consent for the use of tracking cookies for behavioural targeting.

Under the empowerment model, policymakers could also try to nudge Internet users towards disclosing less information. For instance, while marketers tend to argue that somebody gives implied consent if she does not opt out of behavioural targeting, policymakers could require firms to obtain the individual's opt-in consent. But behavioural sciences casts doubt on the potential of opt-in consent as a privacy protection measure. People tend to click OK to almost any request they see pop up on their screens. This is all the more true when firms make the use of a service conditional on the user's consent.

---

[1] The chapter builds on research of my PhD thesis: 'Improving privacy protection in the area of behavioural targeting'. A working paper that forms the basis for the chapter has been presented at the 6th Annual Privacy Law Scholars Conference (Berkeley, 7 June 2013), and the Nudging in Europe conference (Liège, 12-13 December 2013). I am grateful for the comments received there. Furthermore, I thank Alberto Alemanno, Axel Arnbak, Bodó Balázs, Oren Bar-Gill, Christian Handke, Stefan Kulk, Florencia Marotta-Wurgler, Aleecia McDonald, Shara Monteleone, Alessio Pacces, Joost Poort, Anne Lise Sibony, Omer Tene, Nico van Eijk, and Joris van Hoboken.

[2] The distinction between empowerment and protection rules is made to structure the discussion, but it is not a formal legal distinction. See section 5.1.

A second approach focuses on protection of the individual. This protective approach can also be recognised in current law regarding privacy. The Data Protection Directive has elements that aim to protect the individual. For example, even after a firm obtains an individual's consent, data protection law does not allow excessive personal data processing. And data protection law always requires firms to secure the data they process. But enforcing data protection law may not be enough to protect privacy in this area. I argue that, if society is better off when certain behavioural targeting practices do not happen, policymakers should consider banning them.

The chapter is structured as follows. Section 2 introduces the practice of behavioural targeting and related privacy problems. Section 3 discusses the current regulatory regime to protect privacy in this area, and shows informed consent plays a central role in the regime. Section 4 analyses problems with informed consent through the lens of behavioural sciences. The section discusses information asymmetry, transaction costs, and biases that influence people's privacy decisions. Taking the behavioural economic insights into account, section 5 discusses two ways to improve privacy protection in the area of behavioural targeting: empowerment and protection of the individual. Section 6 concludes.

## 2 Behavioural Targeting and Privacy

Much of the commercial data collection on the Internet is driven by behavioural targeting, a type of electronic direct marketing. Vast amounts of information about hundreds of millions of people is collected for behavioural targeting.

In a simplified example, behavioural targeting involves three parties: an Internet user, a website publisher, and an advertising network. Advertising networks are firms that serve ads on thousands of websites, and can recognise people when they browse the web. An ad network might infer that somebody who often visits websites about tennis is a tennis enthusiast. If that person visits a news website, the ad network might display advertising for tennis rackets. When simultaneously visiting that same website, somebody who visits many websites about economics might see ads for economics books.

A commonly used technology for behavioural targeting involves cookies. A cookie is a small text file that a website publisher stores on a user's computer to recognise that device during subsequent visits. Many websites use cookies, for example to remember the contents of a virtual shopping cart (first party cookies). Ad networks can place and read cookies as well (third party cookies). As a result, an ad network can follow an Internet user across all websites on which it serves ads. Third party tracking cookies are placed through virtually every popular website. A visit to one website often leads to receiving third party cookies from dozens of ad networks. In addition to cookies, firms use other tracking technologies for behavioural targeting, such as various kinds of super cookies, device fingerprinting and deep packet inspection. Therefore, deleting cookies is not always enough to prevent being tracked.[3]

Behavioural targeting could benefit firms and consumers. Advertising funds an astonishing amount of Internet services. Without paying with money, people enjoy access to online translation tools, online newspapers, and email accounts, and can watch videos and listen to music. But behavioural targeting also raises privacy concerns.

Surveys show that most people do not want behaviourally targeted advertising, because they find it creepy or privacy-invasive. A small minority indicates it does not mind the data collection and prefers behaviourally targeted advertising because it can lead to more relevant ads.[4]

Three of the main privacy problems regarding behavioural targeting are (i) chilling effects, (ii) a lack of control over personal information, and (iii) the risk of unfair discrimination and manipulation. First, chilling effects can occur because of the massive collection of information about people's online activities. Firms compile detailed profiles based on what Internet users read, what videos they watch, what they search for, etc. People may adapt their behaviour if they suspect their activities may be monitored.[5] For example, somebody who fears surveillance might hesitate to look for medical information on the web, or to read about political topics. Regardless of how data are used at later stages, the mere collection of data can cause a chilling effect.

---

[3] See generally on various tracking technologies: Hoofnagle CJ et al, 'Behavioral Advertising: The Offer You Cannot Refuse' (2012) 6(2) Harvard Law & Policy Review 273, 291; Kuehn A and Mueller M. 'Profiling the profilers: deep packet inspection and behavioral advertising in Europe and the United States' (1 September 2012) <http://ssrn.com/abstract=2014181> accessed on 16 August 2014.

[4] See: Turow J et al, 'Americans Reject Tailored Advertising and Three Activities that Enable it' (29 September 2009) <http://ssrn.com/abstract=1478214> accessed 30 June 2014. In Europe, seven out of ten people are concerned that firms might use data for new purposes such as targeted advertising without informing them (European Commission, 'Special Eurobarometer 359: Attitudes on data protection and electronic identity in the European Union' (2011) <http://ec.europa.eu/public_opinion/archives/ebs/ebs_359_en.pdf> accessed 18 November 2012, p 146. See for an overview of studies on people's attitudes towards behavioural targeting: my papers that are mentioned in footnote 1.

Second, people lack control over data concerning them. People do not know which information is collected, how it is used, and with whom it is shared (see section 4.1). The feeling of lost control is a privacy problem. And large-scale personal data storage brings risks. For instance, a data breach could occur, or data could be used for unexpected purposes, such as identity fraud.[6]

Third, behavioural targeting enables social sorting and discriminatory practices: firms can classify people as 'targets' and 'waste', and treat them accordingly.[7] And some fear that behavioural targeting could be used to manipulate people. Personalised advertising could become so effective that advertisers have an unfair advantage over consumers.[8] Calo warns for 'digital market manipulation' and 'mass production of bias'.[9] With modern personalised marketing techniques, 'companies will be in a

---

[5] See: Foucault M, *Discipline and punish: The birth of the prison (transl. Sheridan A.)* (Random House LLC 1977); Richards NM, 'Intellectual privacy' (2008) 87 Texas Law Review 387.

[6] Gürses S, *Multilateral Privacy Requirements Analysis in Online Social Networks (PhD thesis University of Leuven)* (KU Leuven (academic version) 2010), p 87-91; Calo MR, 'The boundaries of privacy harm' (2011) 86 Indiana Law Journal 1131.

[7] Turow J, *The Daily You: How the New Advertising Industry is Defining Your Identity and Your Worth* (Yale University Press 2011). In Lyon's words, social sorting involves 'obtain[ing] personal and group data in order to classify people and populations according to varying criteria, to determine who should be targeted for special treatment, suspicion, eligibility, inclusion, access, and so on' (Lyon D, 'Surveillance as social sorting: computer codes and mobile bodies ' in Lyon D. (ed), *Surveillance as social sorting: privacy, risk and automated discrimination* (Routledge 2002), p 20).

[8] See on unfair commercial practices and behavioural targeting: European Data Protection Supervisor, 'Privacy and competitiveness in the age of big data: The interplay between data protection, competition law and consumer protection in the Digital Economy (March 2014) <https://secure.edps.europa.eu/EDPSWEB/webdav/site/mySite/shared/Documents/Consultation/Opinions/2014/14-03-26_competitition_law_big_data_EN.pdf> accessed 11 April 2014.

[9] Calo MR. 'Digital market manipulation (George Washington Law Review, Forthcoming; University of Washington School of Law Research Paper No. 2013-27.)' (2013) <http://ssrn.com/abstract=2309703> accessed 16 February 2014, p 12.

position to figure out where and how consumers are irrational, and exploit that irrationality for gain.'[10]

Others worry that excessive personalisation can lead to an 'information cocoon',[11] or a 'filter bubble': 'a unique universe of information for each of us'.[12] This fear seems most relevant when firms personalise not only ads, but also other content and services. Briefly stated, the idea is that personalised advertising and other content could surreptitiously steer people's choices. For example, a search engine provider that personalises search results might provide links to conservative news to somebody who mainly visits conservative news sites. If the user thinks she sees a neutral picture, such personalisation might influence her worldview.

## 3 Informed Consent in the Law

### 3.1 Data Protection Law

The right to privacy is a fundamental right in the European legal system, and is included in the European Convention on Human Rights (1950). The European Court of Human Rights interprets the Convention's privacy right generously, and holds that information derived from monitoring somebody's Internet usage is protected under the right to privacy.[13] The European Union Charter of Fundamental Rights copies the Convention's right to privacy almost verbatim. In addition, the Charter grants individuals a separate right to the protection of personal data.[14]

---

[10] Calo MR, 'Why Opt Out of Tracking? Here's a Reason' (14 August 2013) <https://cyberlaw.stanford.edu/blog/2013/08/why-opt-out-tracking-heres-reason> accessed 13 August 2014.

[11] Sunstein CR, *Infotopia: How many minds produce knowledge* (Oxford University Press 2006), p 9.

[12] Pariser E, *The Filter Bubble* (Penguin Viking 2011), p 9. Some authors are sceptical about the risks of filter bubbles. See e.g. Van Hoboken JVJ, Search engine freedom: on the implications of the right to freedom of expression for the legal governance of search engines (PhD thesis University of Amsterdam) (Information Law Series, Kluwer Law International 2013), p 286-287; p 301.

[13] ECtHR, Copland v. United Kingdom, No. 62617/00, 3 April 2007, par 41-42.

[14] Art 8 of the Charter of Fundamental Rights of the European Union: "Protection of personal data" reads:
"1. Everyone has the right to the protection of personal data concerning him or her.
2. Such data must be processed fairly for specified purposes and on the basis of the consent of the person concerned or some other legitimate basis laid down by law. Everyone has the right of access to To protect privacy in the area of behavioural targeting, the main legal instruments in Europe are the general Data Protection Directive (1995),[15] and the e-Privacy Directive's consent requirement for tracking technologies (2009).[16] Informed consent plays an important role in both directives.

Data protection law is a legal tool that aims to ensure that the processing of personal data happens fairly and transparently.[17] Data protection law grants rights to people whose data are being processed (data subjects), and imposes obligations on parties that process personal data (data controllers, limited to and referred to as firms in this chapter).[18] Independent Data Protection Authorities (hereinafter DPAs) oversee compliance with the rules.[19] European DPAs cooperate in the Article 29 Working Party, an independent advisory body.[20] The Working Party's publishes opinions on the interpretation of data protection law, which, although not legally binding, are influential. Judges and DPAs often follow the Working Party's interpretation.[21]

Since its inception in the early 1970s, data protection law has evolved into a complicated field of law. Borrowing from Bygrave, the core of data protection law can be summarised in nine data protection principles: the principle of fair and lawful processing, the transparency principle, the principle of data subject participation and control, the purpose limitation principle, the data minimisation principle, the

---

data which has been collected concerning him or her, and the right to have it rectified."
3. Compliance with these rules shall be subject to control by an independent authority.

[15] Directive 95/46/EC of the European Parliament and of the Council of 24 October 1995 on the protection of individuals with regard to the processing of personal data and on the free movement of such data [1995] OJ L 281/31.

[16] Directive 2002/58/EC of the European Parliament and of the Council of 12 July 2002 concerning the processing of personal data and the protection of privacy in the electronic communications sector (Directive on privacy and electronic communications) [2002] OJ L 201/37, as amended by Directive 2006/24/EC [the Data Retention Directive], and Directive 2009/136/EC [the Citizen's Rights Directive].

[17] See on data protection law as a "transparency tool": De Hert P and Gutwirth S, 'Privacy, Data Protection and Law Enforcement. Opacity of the Individual and Transparency of Power' in Claes E, Duff A and Gutwirth S (eds), *Privacy and the Criminal Law* (Intersentia 2006).

[18] Art 2(a) and 2(h) of the Data Protection Directive.

[19] Art 8(3) of the European Union Charter of Fundamental Rights.

[20] The Article 29 Working Party is set up by art 29 of the Data Protection Directive.

[21] See: Gutwirth S and Poullet Y, 'The contribution of the Article 29 Working Party to the construction of a harmonised European data protection system: an illustration of 'reflexive governance'?' in Asinari VP and Palazzi P (eds), *Défis du Droit à la Protection de la Vie Privée. Challenges of Privacy and Data Protection Law* (Bruylant 2008).

information quality principle, the proportionality principle, the security principle, and the sensitivity principle.[22]

Data protection law only applies when ‘personal data’ are processed: data that relate to an identifiable person.[23] Behavioural targeting often involves processing pseudonymous data: individual but nameless profiles. Many firms that do behavioural targeting claim they only process ‘anonymous’ data, and that data protection law does not apply to their activities.[24] But the Article 29 Working Party holds that behavioural targeting generally entails the processing of personal data, also if a firm cannot tie a name to the data it has on an individual. If a firm aims to use data to ‘single out’ a person, or to distinguish a person within a group, these data are personal data, according to the Working Party.[25]

In January 2012 the European Commission presented a proposal for a Data Protection Regulation,[26] which should replace the 1995 Data Protection Directive. At the time of writing, it is unclear whether the proposal will be adopted. The most optimistic view seems to be that the Regulation could be adopted in 2015.[27] While based on the same principles as the Directive, the proposal would bring significant changes. For instance, unlike a directive, a regulation has direct effect and does not have to be transposed in the national laws of the member states. Under the proposal, DPAs could impose high fines. And the proposal always requires consent to be 'explicit'.[28] The proposal's preamble emphasises the ideal of data subject control. 'Individuals should have control of their own personal data.'[29]

---

[22] Bygrave LA, *Data privacy law. An international perspective* (Oxford University Press 2014), ch 5. Bygrave discusses 8 principles, and sees the transparency principle as a part of the fair and lawful principle. I use a slightly different terminology than Bygrave.

[23] Art 2(a), 2(b), and 3(1) of the Data Protection Directive.

[24] See for instance the Interactive Advertising Bureau: ‘Data about your browsing activity is collected and analysed anonymously’ (Interactive Advertising Bureau Europe. 'Your Online Choices. A Guide to Online Behavioural Advertising’. FAQ # 22. <www.youronlinechoices.com/ma/faqs#22> accessed 17 February 2014).

[25] Article 29 Working Party, 'Opinion 2/2010 on online behavioural advertising’ (WP 171), 22 June 2010, p 9; Article 29 Working Party, 'Opinion 05/2014 on Anonymisation Techniques’ (WP 216) 10 April 2014.

[26] European Commission, Proposal for a Regulation of the European Parliament and of the Council on the Protection of Individuals with regard to the Processing of Personal Data and on the Free Movement of Such Data (General Data Protection Regulation) COM(2012) 11 final, 2012/0011 (COD), 25 January 2012.

[27] See: European Council, ‘Conclusions of the European Council 26/27 June 2014’ <www.consilium.europa.eu/uedocs/cms_Data/docs/pressdata/en/ec/143478.pdf> accessed 29 June 2014, p 2.

### 3.2 Informed Consent for Personal Data Processing

The current Data Protection Directive only allows firms to process personal data if they can base the processing on consent or on one of five other legal grounds. For the private sector, the most relevant legal grounds are: a contract, the balancing provision, and the data subject's consent.[30]

A firm can process personal data if the processing is necessary for the performance of a contract with the data subject.[31] For example, certain data have to be processed for a credit card payment, or for a newspaper subscription. The 'necessary' requirement sets a higher threshold than 'useful' or 'profitable'.[32] Some Internet firms suggest a user enters a contract by using their services, and that it is necessary for this contract to track the user for behavioural targeting.[33] However, according to the Working Party, a firm can only rely on this legal ground if the processing is genuinely necessary for the provision of the service. The Working Party says that a contract is not a suitable legal ground for the processing of personal data for behavioural targeting.[34] In any case, the practical problems with informed consent to behavioural targeting which are discussed below would be largely the same if firms could base the processing for behavioural targeting on a contract.

---

[28] See on consent: art 4(8) and art 7; on fines: art 78-79 of the European Commission proposal for a Data Protection Regulation (2012).
[29] Recital 6 of the European Commission proposal for a Data Protection Regulation (2012).
[30] The legal grounds are listed in art 7 of the Data Protection Directive.
[31] Art 7(b) of the Data Protection Directive.
[32] See: ECtHR, Silver and Others v. United Kingdom (App 5947/72) (1983) 5 EHRR 347, par 97: "The adjective 'necessary' is not synonymous with 'indispensable', neither has it the flexibility of such expressions as 'admissible', 'ordinary', 'useful', 'reasonable' or 'desirable' (...)."
[33] See e.g. College Bescherming Persoonsgegevens, 'Investigation into the combining of personal data by Google, Report of Definitive Findings' (z2013-00194) (November 2013, with correction 25 November 2013) <www.dutchdpa.nl/downloads_overig/en_rap_2013-google-privacypolicy.pdf> accessed 1 May 2014, p 11.
[34] Article 29 Working Party, 'Opinion 06/2014 on the notion of legitimate interests of the data controller under article 7 of Directive 95/46/EC' (WP 217) 9 April 2014, p 16-17.

The balancing provision allows certain data processing activities on an opt-out basis, without the data subject's prior consent. The balancing provision allows processing when it is necessary for the firm's legitimate interests, except where such interests are overridden by the data subject's interests or fundamental rights.[35] The balancing provision is the appropriate legal ground for innocuous standard business practices. For example, a firm can generally rely on the balancing provision for postal direct marketing of its own products to current or past customers. If a firm relies on the balancing provision for direct marketing, data protection law grants the data subject the right to stop the processing, by opting out.[36] The Data Protection Directive does not say explicitly whether behavioural targeting can be based on the balancing provision. But the most convincing view is that firms cannot base personal data processing for behavioural targeting on the balancing provision, in particular when it involves tracking somebody over multiple websites.[37] The data subject's interests must generally prevail over the firm's business interests, as behavioural targeting involves collecting and processing information about personal matters such as people's browsing behaviour. The Working Party confirms that firms can almost never rely on the balancing provision to process personal data for behavioural targeting.[38]

If firms want to process personal data, and cannot base the processing on the balancing provision or another legal ground, they must ask the data subject for consent.[39] The Working Party says consent is generally the required legal ground for personal data processing for behavioural targeting.[40] It follows from the Data Protection Directive's consent definition that consent requires a free, specific, and informed indication of wishes.[41] People can express their will in any form, but mere silence or inactivity is not an expression of will. This is also the predominant view in general contract law.[42] During the drafting of the Directive in the early 1990s, firms claimed the law should allow them to obtain 'implied' consent by offering people the chance to object, with an opt-out system.[43] Presumably firms assumed it would be easier to obtain consent through opt-out systems than through opt-in systems. But the European policymaker rejected the idea that somebody expresses her will to consent to processing, merely because she fails to object.[44]

---

[35] Art 7(f) of the Data Protection Directive.
[36] Art 14(b) of the Data Protection Directive.
[37] See for instance: Traung P, 'EU Law on Spyware, Web Bugs, Cookies, etc. Revisited: Article 5 of the Directive on Privacy and Electronic Communications' (2010) 31 Business Law Review 216; Van Der Sloot B, 'Het plaatsen van cookies ten behoeve van behavioural targeting vanuit privacyperspectief [The placing of cookies for behavioural targeting from a privacy perspective]' (2011)(2) Privacy & Informatie 62.
[38] Article 29 Working Party, 'Opinion 06/2014 on the notion of legitimate interests of the data controller under art 7 of Directive 95/46/EC' (WP 217) 9 April 2014, p 45.
[39] Art 7(a) of the Data Protection Directive.
[40] Article 29 Working Party, 'Opinion 06/2014 on the notion of legitimate interests of the data controller under art 7 of Directive 95/46/EC' (WP 217) 9 April 2014, p 45.

In line with the transparency principle, consent has to be specific and informed. Consent cannot be valid if a consent request does not include a specified processing purpose and other information that is necessary to guarantee fair processing.

Furthermore, consent must be 'free'. For instance, if an employer asks an employee for consent, the consent might not be sufficiently voluntary, because of the imbalance of power. The employee might fear adverse consequences if she does not consent.[45] Negative pressure would make consent invalid, but positive pressure is generally allowed.[46] In many circumstances, current data protection law probably allows firms to offer take-it-or-leave-it choices. One data protection handbook contends: '[i]f, for instance, not consenting to having a supermarket's customer card results only in not receiving deductions from prices of certain goods, consent is still a valid legal basis

---

[41] Art 2(h) of the Data Protection Directive.
[42] 'A statement made by or other conduct of the offeree indicating assent to an offer is an acceptance. Silence or inactivity does not in itself amount to acceptance', says art 18(1) of the Vienna Convention on International Sale of Goods for instance. The European Court of Justice affirms that consent cannot be inferred from inactivity. See for instance: CJEU, Cases C-92/09 and C-93/09, Schecke and Eifert [2010] ECR 1-11063, par 63.
[43] See for instance: International Chamber of Commerce, 'Protection of Personal Data: an International Business View' (1992)(8) Computer Law and Security Report 259, p. 262.
[44] Kosta E, *Consent in European Data Protection Law (PhD thesis University of Leuven)* (Martinus Nijhoff Publishers 2013), p 83-108.
[45] Article 29 Working Party, 'Opinion 15/2011 on the definition of consent' (WP 187) 13 July 2011, p 13-14.
[46] Kosta E, *Consent in European Data Protection Law (PhD thesis University of Leuven)* (Martinus Nijhoff Publishers 2013), p 256.

for processing personal data of those customers who consented to having such a card.'[47]

In sum, behavioural targeting generally entails the processing of personal data, also if firms do not tie a name to the data they have on individuals. The law usually requires firms to obtain the data subject's informed consent for personal data processing for behavioural targeting.

### 3.3 Informed Consent for Tracking Technologies

European legal discussions on behavioural targeting often focus on the e-Privacy Directive's consent requirement for tracking technologies, rather than on the general data protection rules. The e-Privacy Directive complements the general data protection regime with more specific privacy rules for the electronic communications sector.

In early proposals for the 2002 version of the e-Privacy Directive, article 5(3) required firms to ask consent before they placed certain kinds of cookies. After fierce lobbying by the marketing industry, the final version used ambiguous wording about a 'right to refuse'. The 2002 version of article 5(3) is usually interpreted as an opt-out system: websites had to offer people the possibility to object to tracking cookies.[48]

The e-Privacy Directive was updated in 2009.[49] Since 2009, article 5(3) requires any party that stores or accesses information on a user's device to obtain the user's informed consent. For the definition of consent, the e-Privacy Directive refers to the Data Protection Directive, which requires a free, specific, and informed expression of will.[50] There are exceptions to the consent requirement, for example for cookies that are strictly necessary for a service requested by the user, and for cookies that are necessary for the transmission of communication. Hence, no prior consent is needed for cookies that are used for a digital shopping cart, or for log-in procedures. Article 5(3) applies regardless of whether personal data are processed. For ease of reading this chapter speaks of consent for 'cookies' or for 'tracking technologies', but article 5(3) applies to any information that can be stored or accessed on a user's device.

---

[47] European Agency for Fundamental Rights, *Handbook on European data protection law* (Publications Office of the European Union 2014), p 59.

[48] Kierkegaard S, 'How the cookies (almost) crumbled: privacy & lobbyism' (2005) 21(4) Computer Law & Security Review 310.

[49] The e-Privacy Directive 2002/58 was updated by Directive 2009/136. This chapter refers to the consolidated version from 2009.

[50] Art 2(f) and recital 17 of the e-Privacy Directive.

As many Member States missed the 2011 implementation deadline of the e-Privacy Directive, most national implementation laws are rather new. Member States vary in their approaches. For example, the Netherlands requires, in short, opt-in consent for tracking cookies.[51] In contrast, the United Kingdom appears to allow firms to use opt-out systems to obtain 'implied' consent.[52] However, the Working Party insists that the data subject's inactivity does not signify consent.[53]

In conclusion, the European legal regime regarding privacy in the area of behavioural targeting focuses heavily on empowering individuals to make choices in their own best interests. Data protection law is deeply influenced by the idea of privacy as control over personal information. This privacy concept is prevalent since 1967, when Westin defined privacy as 'the claim of individuals, groups or institutions to determine when, how and to what extent information about them is communicated to others.'[54] Proposals to reform European data protection law also emphasise giving people control over personal information concerning them.

## 4 Informed Consent in Practice

People's choices regarding privacy can be analysed using economic theory. Acquisti, one of the leading scholars on the economics of privacy, explains: 'privacy economics deals with informational trade-offs: it tries to understand, and sometimes quantify, the costs and benefits that data subjects (as well as potential data holders) bear or enjoy when their personal information is either protected or shared.'[55] Through an economic lens, consent to behavioural targeting can be seen as a transaction between a person and a firm. The user discloses personal information, in exchange for the use of a so-called 'free' service. But this 'transaction' is plagued by market failures. This chapter focuses on the market failure of information asymmetry.[56] Furthermore, there are 'behavioural market failures' regarding online privacy.[57]

---

[51] The Dutch situation is a bit more complicated: see: Zuiderveen Borgesius FJ, 'Behavioral Targeting. Legal Developments in Europe and the Netherlands' (position paper for W3C Workshop: Do Not Track and Beyond), 2012, <www.w3.org/2012/dnt-ws/position-papers/24.pdf> accessed 30 June 2014.
[52] Information Commissioner's Office, 'Changes to cookies on our website', 31 January 2013, <www.ico.org.uk/news/current_topics/changes-to-cookies-on-our-website> accessed 30 June 2014.
[53] Article 29 Working Party, 'Working Document 02/2013 providing guidance on obtaining consent for cookies' (WP 208) 2 October 2013.
[54] Westin AF, *Privacy and Freedom* (The Bodley Head 1970 (1967)), p 7. See for alternative privacy concepts: Gürses S, *Multilateral Privacy Requirements Analysis in Online Social Networks (PhD thesis University of Leuven)* (KU Leuven (academic version) 2010).

## 4.1 Information Asymmetries

Information asymmetry 'means that a party in an interaction may know more about the activity that it is engaged in than does the other party.'[58] Since the 1970s economists devote much attention to markets with asymmetric information, for example where consumers have difficulties evaluating the quality of products or services. Akerlof used the market for used cars as an example.[59] Suppose sellers offer good cars and bad cars ('lemons'). Sellers know whether they have a good car or a lemon for sale, but buyers cannot detect hidden defects. A rational buyer will offer the price corresponding to the average quality of all used cars on the market. But this means that sellers of good cars are offered a price that is too low. Hence, owners of good cars will not offer their cars for sale. The result is that the average quality of used cars on the market decreases. Buyers will therefore offer lower prices, and fewer people will offer their cars for sale. The average quality of cars on the market will drop. Sellers thus do not compete on quality in a market characterised by asymmetric information about quality. Such a lemons situation can lead to products or services of low quality: a race to the bottom.

---

[55] Acquisti A, 'Nudging Privacy. The behavioral economics of personal information' (2009) 7(6) IEEE Security & Privacy 82, p 72.

[56] Other market failures are also relevant for behavioural targeting, such as market power and externalities. See my papers that are mentioned in footnote 1.

[57] The phrase 'behavioural market failures' was introduced by Bar-Gill O, *Seduction by Contract: Law, Economics, and Psychology in Consumer Markets* (Oxford University Press 2012).

[58] Chang HJ, *Economics: the user's guide (paperback)* (Penguin 2014), p 391.

[59] Akerlof GA, 'The Market for "Lemons": Quality Uncertainty and the Market Mechanism' (1970) 84(3) Quarterly Journal of Economics 488.

Information asymmetry is a type of market failure, which, from an economic perspective, justifies regulatory intervention, provided that legal intervention does not bring too much costs or economic distortions.[60] Following this line of reasoning, the main reason for responding to information asymmetry is protecting a well-functioning market, rather than paternalistic motives towards the consumer.

The current state of affairs regarding behavioural targeting is characterised by large information asymmetries. To make an informed choice, people must realise they make a choice. But research shows that most people are only vaguely aware that data are collected for behavioural targeting. For example, Ur et al found in interviews that participants were 'surprised to learn that browsing history is currently used to tailor advertisements'.[61] In a survey by Cranor & McDonald only 40% of respondents thought that providers of email services scan the contents of messages to serve targeted advertising. 29% thought this would never happen, either because the law prohibits it, or because the consumer backlash would be too great. Cranor & McDonald conclude that people generally lack the knowledge needed to make meaningful decisions about privacy in the area of behavioural targeting.[62] In addition, people who have learned how to defend themselves against tracking must update their knowledge constantly. For example, many firms have used flash cookies to respawn (re-install) cookies that people deleted. Hoofnagle et al summarise: 'advertisers are making it impossible to avoid online tracking.'[63]

---

[60] This chapter speaks of economics for ease of reading, but it would be more correct to speak of 'neoclassical economics', as there are also other schools of economic thought. See Chang HJ, *Economics: the user's guide (paperback)* (Penguin 2014), p 109-169.

[61] Ur B et al, 'Smart, Useful, Scary, Creepy: Perceptions of Online Behavioral Advertising' (Proceedings of the Eighth Symposium on Usable Privacy and Security ACM, 2012) 4, p 4. The interviews were conducted in the United States. There is little evidence that in Europe in 2014 people have a better understanding of behavioural targeting.

[62] Cranor LF and McDonald AM. 'Beliefs and Behaviors: Internet Users' Understanding of Behavioral Advertising (38th Research Conference on Communication, Information and Internet Policy (Telecommunications Policy Research Conference)) (2 October 2010) <http://ssrn.com/abstract=1989092> accessed 30 June 2014. The research was conducted in the United States.

[63] Hoofnagle CJ et al, 'Behavioral Advertising: The Offer You Cannot Refuse' (2012) 6(2) Harvard Law & Policy Review 273, p 273.

As Acquisti notes, if firms did ask consent for behavioural targeting, information asymmetry would still be a problem.[64] First, there are many firms involved in serving behaviourally targeted ads, and the underlying data flows are complicated. In addition, people do not know what will happen with their data. Will their name be tied to the profile of their surfing behaviour? Will their data be shared with other firms? If a firm goes bankrupt, will its database be sold to the highest bidder? Second, even if people knew what firms did with their data, it would be hard to predict the consequences. If a firm shares data with another firm, will the data be used for price discrimination? Will visits to a website with medical information lead to higher health insurance costs? If there's a data breach at a firm, will this lead to identity fraud?

If the privacy-friendliness of websites is seen as a product feature, the web has characteristics of a lemons market, say Vila et al.[65] It is hard for people to determine how much of their personal information is captured during a website visit, and how the information will be used. Some firms try to compete on privacy. To illustrate, a couple of search engine providers advertise they do not collect user data.[66] But website publishers rarely use privacy as a competitive advantage. Virtually every popular website allows third parties to track its visitors.[67] 'This situation looks like the classic market for lemons problem', says Strandburg. 'Consumers cannot recognize quality (here, absence of data collection for advertising) and hence will not pay for it. As a result, the market spirals downward.'[68] After interviewing people in the online marketing business, Turow concludes that competition pushes firms towards privacy invasive marketing practices, which confirms the lemons situation.[69] There also seems

[64] See: Acquisti A and Grossklags J, 'What Can Behavioral Economics Teach Us About Privacy?' in Acquisti A et al (eds), *Digital Privacy: Theory, Technologies and Practices* (Auerbach Publications, Taylor and Francis Group 2007).

[65] Vila T, Greenstadt R and Molnar D, 'Why We Can't be Bothered to Read Privacy Policies. Models of Privacy Economics as a Lemons Market.' in Camp, L. J. and S. Lewis (eds), *Economics of Information Security* (Springer 2004).

[66] Two examples are: <www.duckduckgo.com> and my own favourite: <www.startpage.com> accessed 18 August 2014.

[67] Hoofnagle, CJ and Good N. 'The web privacy census' (October 2012) <http://law.berkeley.edu/privacycensus.htm> accessed 30 June 2014.

[68] Strandburg KJ, 'Free Fall: the Online Market's Consumer Preference Disconnect' (2013) University of Chicago Legal Forum 95, p 156.

[69] Turow J, *The Daily You: How the New Advertising Industry is Defining Your Identity and Your Worth* (Yale University Press 2011), p 199.

to be a lemons problem in the market for smartphone applications and social network sites.[70]

## 4.2 Transaction Costs

Data protection law aims to reduce the information asymmetry by requiring firms to disclose certain information to data subjects. The Data Protection Directive obliges firms to provide the data subject information about their identity and the processing purpose, and all other information that is necessary to guarantee fair processing.[71] Website publishers can use a privacy policy to comply with data protection law's transparency requirements. These requirements also apply if a firm does not seek the data subject's consent, but relies on another legal ground for data processing.

But almost nobody reads privacy policies or consent requests. To illustrate, an English computer game store obtained the soul of 7500 people. According to the website's terms and conditions, customers granted 'a non transferable option to claim, for now and for ever more, your immortal soul,' unless they opted out. By opting out, people could save their soul, and could receive a 5 pound voucher. But few people opted out. The firm later said it would not exercise its rights.[72]

Marotta-Wurgler did research on the readership of end-user license agreements (EULAs) of software products. She analysed the click streams of almost 50.000 households, and finds an 'average rate of readership of EULAs (…) on the order of 0.1 percent to 1 percent.' On average, those readers did not look long enough at EULA to read them.[73] 'The general conclusion is clear: no matter how prominently

---

[70] See on social network sites and information asymmetry: Bonneau J and Preibusch S, 'The privacy jungle: On the market for data protection in social networks' in Moore T, Pym DJ and Loannidis C (eds), *Economics of information security and privacy* (Springer 2010).
[71] Art 10 and 11 of the Data Protection Directive.
[72] Fox News, '7,500 Online Shoppers Unknowingly Sold Their Souls' (15 April 2010) <www.foxnews.com/tech/2010/04/15/online-shoppers-unknowingly-sold-souls/> accessed 30 June 2014.
[73] Marotta-Wurgler F, 'Will Increased Disclosure Help? Evaluating the Recommendations of the ALI's Principles of the Law of Software Contracts' (2011) 78(1) The University of Chicago Law Review 165, p 168

EULAs are disclosed, they are almost always ignored.'[74] There is little reason to assume the readership of privacy policies is much higher.

It is not surprising that privacy policies are hardly read: the transaction costs are too high.[75] Reading privacy policies would cost too much time, as they're often long, difficult to read, and vague. Cranor & McDonald calculate that it would cost someone several weeks per year to read the privacy policies of the websites she visits. The opportunity costs to citizens to inform themselves exceeded the revenues from the ad industry they might try to protect themselves from.[76] Furthermore, the language in privacy policies is too difficult for many. (I often have trouble deducing from a privacy policy what a firm plans to do with personal data.) A quarter of Europeans say privacy policies are too difficult.[77] In one study, more than half of the examined privacy policies was too difficult for a majority of American Internet users.[78]

If somebody read and understood a privacy policy, transaction costs could still be a problem. Moving to another service often involves transaction costs for the user, and 'when the costs of switching from one brand of technology to another are substantial, users face *lock-in*'.[79] For instance, transferring emails and contacts to another email provider costs time. If iTunes changes its privacy policy, many people might just accept. And when all one's friends are on Facebook, it makes little sense to join another social network site. Firms can also use transaction costs strategically, for example to discourage people from opting out of tracking. To illustrate, opting out of receiving Google's advertising cookies takes five mouse clicks from its search page.[80]

---

[74] Ibid, p 182.

[75] Transaction costs are 'any costs connected with the creation of transactions themselves, apart from the price of the good that is the object of the transaction' (Luth HA, *Behavioural Economics in Consumer Policy: The Economic Analysis of Standard Terms in Consumer Contracts Revisited (PhD thesis University of Rotterdam),* Intersentia 2010, p 19). See also: Coase RH, 'The Problem of Social Cost' (1960) 3 Journal of Law and Economics 1.

[76] Expressed in dollars, the opportunity cost of reading would be around 781 billion dollars. All online advertising income in the United States was estimated to be 21 billion dollar in 2007. McDonald AM and Cranor LF, 'The Cost of Reading Privacy Policies' (2008) 4(3) I/S: A Journal of Law and Policy for the Information Society 540.

[77] European Commission, 'Special Eurobarometer 359: Attitudes on data protection and electronic identity in the European Union' (2011) <http://ec.europa.eu/public_opinion/archives/ebs/ebs_359_en.pdf> accessed 30 June 2014, p 112-114.

[78] Jensen C and Potts C, 'Privacy policies as decision-making tools: an evaluation of online privacy notices' (2004), Proceedings of the SIGCHI Conference on Human Factors in Computing Systems 471.

[79] Shapiro C and Varian HR, *Information Rules. A Strategic Guide to the Network Economy* (Harvard Business School Press 1999), p 104.

Furthermore, it is often questionable whether people have a real choice regarding tracking. For instance, an Internet firm might have a dominant position or a monopoly.[81] Moreover, even if the market structure is competitive (in other words: when no firm has a monopoly or a similar position), there might not be any privacy-friendly competitors. After all, in a market with information asymmetry firms will generally not compete on quality or, in this case, on privacy-friendliness. Besides, somebody who wants to visit website X, may not see website Y as a valid alternative. As Helberger puts it, "media is speech, and when consuming media content it does matter who the speaker is."[82]

Privacy policies thus fail to inform people that use computers. It is even harder to inform people who use mobile devices with smaller screens. In sum, data protection law does not solve the information asymmetry problem.

### 4.3 Status Quo Bias

A hypothetical fully rational person would know how to deal with information asymmetry and uncertainty. For instance, she could base her decision on what happens to people's personal data on average, and she would not be optimistic about quality in a market with information asymmetry. But behavioural sciences insights suggest that people do not tend to deal with information asymmetry in a 'rational' way. Rather, they often rely on rules of thumb, or heuristics. Usually such mental shortcuts work fine, but they can also lead to behaviour that is not in people's self-interest. Privacy choices are influenced by several biases, such as the status quo bias and myopia.

---

[80] See: College bescherming persoonsgegevens (Dutch DPA), 'Investigation into the combining of personal data by Google, Report of Definitive Findings' (z2013-00194) <www.dutchdpa.nl/downloads_overig/en_rap_2013-google-privacypolicy.pdf> accessed 30 June 2014, p 82. See generally on tactics that a firm can use to keep people in the default position that it prefers: Willis LE, 'Why Not Privacy by Default?' (2013) 29 Berkeley Technology Law Journal 61.

[81] See on competition law and personal data: European Data Protection Supervisor, 'Privacy and competitiveness in the age of big data: The interplay between data protection, competition law and consumer protection in the Digital Economy (March 2014) <https://secure.edps.europa.eu/EDPSWEB/webdav/site/mySite/shared/Documents/Consultation/Opinions/2014/14-03-26_competitition_law_big_data_EN.pdf> accessed 11 April 2014.

[82] See: Helberger N, 'Freedom of Expression and the Dutch Cookie-Wall' (2013), Institute for Information Law <www.ivir.nl/publications/helberger/Paper_Freedom_of_expression.pdf> accessed 18 August 2014, p 12.

The status quo bias, or inertia, refers to the power of the default.[83] Most people do not change the default option. This means that the default setting will have a big impact on the dynamics between the firm and the users. A famous example of the status quo bias concerns the percentage of organ donors. Countries that use an opt-out system (people donate their organs unless they express that they do not want to donate) have many donors, while countries that use an opt-in system have few donors.[84]

Marketers can leverage the status quo bias. Free trial periods of newspapers can lead to subscriptions for years, because – in line with the status quo bias – people do not get around to cancelling. 'Buy this pack of shampoo, and get a 2 euro refund', relies on transaction costs and the status quo bias. With such mail-in-rebates, many people fail to to send in the coupon. As an aside, customers would also provide the firm with personal data, such as their name and bank account number, if they sent in the coupon.

Insights into the status quo bias help to understand the discussion about opt-in versus opt-out systems for behavioural targeting and other types of direct marketing. This opt-in/opt-out discussion basically concerns the question of who benefits from the status quo bias: the firm or the individual? As Sunstein puts it, 'true, we might opt out of a website policy that authorizes a lot of tracking (perhaps with a simple click) – but because of the power of inertia, many of us are not likely to do so.'[85] Marketers tend to argue they can obtain 'implied' consent with opt-out systems. Privacy advocates tend to prefer opt-in systems for privacy-intrusive practices.

---

[83] Samuelson W and Zeckhauser R, 'Status Quo Bias in Decision Making' (1988) 1(1) Journal of Risk and Uncertainty 7.
[84] Johnson EJ and Goldstein D, 'Do Defaults Save Lives?' (2003) 302(5649) Science 1338. For a discussion of nudging in this context, see in this volume the chapters by Muireann Quigly and Elen Strokes and that by Alberto Alemanno.
[85] Sunstein CR, 'The Storrs Lectures: Behavioral Economics and Paternalism' (2013) 122(7) Yale Law Journal 1826, p 1893.

The opt-in versus opt-out discussion has been going on for decades. When the European Commission presented a proposal for a Data Protection Directive in 1990, heated discussions ensued about the proposal's rules on direct marketing. Business organisations, including the European Direct Marketing Association, started lobbying intensely.[86] Marketers feared that direct mail marketing would only be allowed with the data subject's prior consent. The lobbying paid off. In 1992 the Commission presented an amended proposal, and said that firms can rely on the balancing provision for direct mail marketing, which implies an opt-out system.[87]

When the 2009 e-Privacy Directive required informed consent for tracking cookies, again an opt-in/opt-out discussion followed. Recital 66 of the 2009 directive that amended the e-Privacy Directive has caused much confusion: 'in accordance with the relevant provisions of [the Data Protection Directive], the user's consent to processing may be expressed by using the appropriate settings of a browser or other application.'[88] Many marketers suggest that people that do not block tracking cookies in their browser give implied consent to behavioural targeting. For instance, the Interactive Advertising Bureau, a marketing trade organisation, says '*default* web browser settings can amount to "consent"'.[89] However, the Working Party insists that the mere fact that somebody leaves the settings of her browser untouched does not mean she expresses her will to accept tracking cookies.[90]

A number of larger behavioural targeting firms offer people the chance to opt out of targeted advertising on a centralised website: youronlinechoices.com. However, participating firms merely promise to stop showing targeted ads, so they may continue to track people who have opted out.[91] In short, the website offers the equivalent of 'do not target', rather than 'do not collect'.[92] But even if the firms stopped collecting data after somebody opts out, they could not use the website's opt-out system to obtain valid consent, says the Working Party.[93] Valid consent requires an expression of will, which generally calls for an opt-in procedure.

---

[86] Regan PM, 'The Globalization of Privacy: Implications of Recent Changes in Europe' (1993) 52(3) American Journal of Economics and Sociology 257, p 266-267; Heisenberg D, *Negotiating Privacy: The European Union, the United States, and Personal Data Protection* (Lynne Rienner Publishers 2005), p 62.

[87] European Commission, amended proposal for a Council Directive on the Protection of Individuals with regard to the Processing of Personal Data and on the Free Movement of Such Data, COM (92) 422 final – SYN 287, 15 October 1992 [1992] OJ C311/30 (27.11.1992), p 15.

[88] Directive 2009/136, recital 66.

[89] Interactive Advertising Bureau United Kingdom, 'Department for Business, Innovation & Skills consultation on implementing the revised EU electronic communications framework, IAB UK Response' (1 December 2012) <www.iabuk.net/sites/default/files/IABUKresponsetoBISconsultationonimplementingtherevisedEUElectronicCommunicationsFramework_7427_0.pdf> accessed 30 June 2014, p 2, emphasis original.

[90] Article 29 Working Party, 'Opinion 2/2010 on online behavioural advertising' (WP 171), 22 June 2010.

The 2012 proposal for a Data Protection Regulation reaffirms that consent must be expressed 'either by a statement or by a clear affirmative action.'[94] History repeated itself: many firms lobbied to soften the requirements for consent. Those firms prefer a regime that allows them to collect personal data, unless people opt out.[95] In part, the lobbying succeeded. The European Parliament amended the proposal for the Regulation. The amended proposal allows firms, under certain conditions, to do behavioural targeting without the data subject's consent, on an opt-out basis (based on the balancing provision), as long as they do not tie a name to the data they process about individuals.[96]

## 4.4 Myopia and Other Biases

More biases influence people's decisions regarding behavioural targeting. For instance, myopia, or present bias, refers to the effect that people tend to focus more on the present than on the future. People often choose for immediate gratification, thereby ignoring future costs. For example, myopia helps to explain why many people find it hard to save money for their retirement.

---

[91] The opt-out page says: 'Declining behavioural advertising only means that you will not receive more display advertising customised in this way' (Youronlinechoices, FAQ# 21 <www.youronlinechoices.com/ma/faqs#21> accessed 30 June 2014).

[92] See on this difference: Schunter M and Swire P. 'Explanatory Memorandum for Working Group Decision on "What Base Text to Use for the Do Not Track Compliance Specification"' (16 July 2013) <www.w3.org/2011/tracking-protection/2013-july-explanatory-memo/> accessed 16 February 2014.

[93] Article 29 Working Party, 'Opinion 16/2011 on EASA/IAB Best Practice Recommendation on Online Behavioural Advertising' (WP 188) 8 December 2011.

[94] Art 4(8) of the European Commission proposal for a Data Protection Regulation (2012); see section 3.1.

[95] See for instance: Facebook, 'Facebook recommendations on the Internal Market and Consumer Affairs draft opinion on the European Commission's proposal for a General Data Protection Regulation 'on the protection of individuals with regard to the processing of personal data and on the free movement of such data'' <https://github.com/lobbyplag/lobbyplag-data/raw/master/raw/lobby-documents/Facebook.pdf> accessed 26 May 2014.

[96] See art 2(a), art 6(f), and recitals 38 and 58a of the proposal for a Data Protection Regulation, consolidated version after LIBE Committee vote, 22 October 2013, <www.janalbrecht.eu/fileadmin/material/Dokumente/DPR-Regulation-inofficial-consolidated-LIBE.pdf> accessed 30 June 2014.

Because of myopia, people might choose immediate access to a service, also if this means they have to consent to behavioural targeting, contrary to their earlier plans. Suppose Alice reads about behavioural targeting, and decides not to accept any more tracking cookies. That night, she wants to read an online newspaper, and wants to watch the news online. Both websites deny entrance to visitors that do not accept the tracking cookies of third parties.[97] While she was planning not to accept any more tracking cookies, Alice clicks 'yes' on both websites.[98]

As noted, under the Data Protection Directive consent must be 'free' to be valid.[99] But in most circumstances data protection law probably allows firms to offer take-it-or-leave-it-choices. In principle, firms are allowed to offer people that consent something in return, such as a discount. This suggests that firms can make the use of a service dependent on consent to behavioural targeting. Hence, in principle current law seems to allow website publishers to install 'tracking walls', barriers that website visitors can only pass if they consent to being tracked.[100]

However, a tracking wall could make consent involuntary if people must use a website. For instance, say people are required to file their taxes online. If the tax website had a tracking wall that imposes third party tracking, people's consent to tracking would not be voluntary. According to the Dutch DPA, the national public broadcasting organisation is not allowed to use a tracking wall, because the only way to access certain information online is through the broadcaster's website.[101] The Working Party emphasises that consent should be free, but does not say that current data protection law prohibits tracking walls in all circumstances.[102]

---

[97] Early 2013 this was the case in the Netherlands. The National Public Broadcasting Organisation and one of the larger newspapers (Volkskrant) both installed a cookie wall (<www.publiekeomroep.nl> and <www.volkskrant.nl> accessed 15 February 2013).
[98] In one Dutch survey, 30% does not want tracking cookies at all, and 41% only wants tracking cookies from some sites. However, 50% say they usually click 'OK' to consent requests for cookies (Consumentenbond, 'Cookiewet heeft bar weinig opgeleverd' [Cookie law didn't help much] (2014) <www.consumentenbond.nl/test/elektronica-communicatie/veilig-online/privacy-op-internet/extra/cookiewet-heeft-weinig-opgeleverd/> accessed 30 June 2014).
[99] Art 2(h) of the Data Protection Directive.
[100] See: Helberger N, 'Freedom of Expression and the Dutch Cookie-Wall' (2013), Institute for Information Law <www.ivir.nl/publications/helberger/Paper_Freedom_of_expression.pdf> accessed 18 August 2014.

Overconfidence and optimism biases are related to myopia. People tend to underestimate the risk of accidents and diseases, and overestimate the chances of a long and healthy life or winning the lottery. The success of 'buy now, pay later' deals can be partly explained by myopia and optimism bias.[103] Research suggests people also tend to underestimate the risks of identity fraud and of re-identification of anonymised data.[104]

The way information is presented can also influence decisions: the framing effect. For example, many people see a link to a privacy policy as a quality seal. 41% of Europeans do not read privacy policies, because they think it is enough to check whether a website has one.[105] In a California survey the majority thought that the mere fact that a website had a privacy policy meant that their privacy was protected by law.[106] Turow at al. argue that the phrase 'privacy policy' is misleading.[107] Facebook speaks of a 'data use policy', which seems a more apt name.[108]

Research suggests that privacy policies with vague language give people the impression that a service is more privacy-friendly than privacy policies that give more details.[109] Another study concludes that 'any official-looking graphic' can lead people to believe that a website is trustworthy.[110] Böhme and Köpsell find that people are more likely to consent if a pop-up looks more like an end user license agreement (EULA). The researchers varied the design of consent dialog boxes and tested the effect by analysing the clicks of more than 80.000 people. They conclude that people are conditioned to click 'agree' to a consent request if it resembles a EULA.

---

[101] College bescherming persoonsgegevens 2013, 'Brief aan de staatssecretaris van Onderwijs, Cultuur en Wetenschap, over beantwoording Kamervragen i.v.m. cookiebeleid' [Letter to the State Secretary of Education, Culture and Science, on answers to parliamentary questions about cookie policy] (31 January 2013) <www.cbpweb.nl/downloads_med/med_20130205-cookies-npo.pdf> accessed 30 June 2014.

[102] Article 29 Working Party 2013, 'Working Document 02/2013 providing guidance on obtaining consent for cookies' (WP 208) 2 October 2013.

[103] Sunstein CR and Thaler RH, Nudge: Improving Decisions about Health, Wealth, and Happiness (Yale University Press 2008), p 31-35.

[104] Acquisti A and Grossklags J, 'Privacy and rationality in individual decision making' (2005) 3(1) IEEE Security & Privacy 26.

[105] European Commission, 'Special Eurobarometer 359: Attitudes on data protection and electronic identity in the European Union' (2011) <http://ec.europa.eu/public_opinion/archives/ebs/ebs_359_en.pdf> accessed 30 June 2014, p 118-120.

[106] Hoofnagle, CJ and King J, 'What Californians Understand about Privacy Online (UC Berkeley)' (3 September 2008) <http://ssrn.com/abstract=1262130> accessed 5 April 2013;

[107] Turow J et al, 'The Federal Trade Commission and Consumer Privacy in the Coming Decade' (2007) 3(3) I/S: A Journal of Law & Policy for the Information Society 723.

[108] Facebook. 'Data Use Policy' (15 November 2013) <www.facebook.com/about/privacy> accessed 30 June 2014.

> [U]biquitous EULAs have trained even privacy-concerned users to click on 'accept' whenever they face an interception that reminds them of a EULA. This behaviour thwarts the very intention of informed consent. So we are facing the dilemma that the long-term effect of well-meant measures goes in the opposite direction: rather than attention and choice, users exhibit ignorance.[111]

Furthermore, Acquisti et al discuss a control paradox. People share more information if they *feel* they have more control over how they share personal information. The researchers conclude that control over personal information is a normative definition of privacy: control *should* ensure privacy. But in practice, '"more" control can sometimes lead to "less" privacy in the sense of higher objective risks associated with the disclosure of personal information.'[112]

### 4.5 Behavioural Market Failures

Behavioural sciences insights can help to explain the alleged privacy paradox. People say in surveys they care about privacy, but often divulge personal data in exchange for minimal benefits. Part of this is conditioning: many people click 'yes' to any statement that is presented to them. It is only a slight exaggeration to say: people do not read privacy policies; if they were to read, they would not understand; if they understood, they would not act.

---

[109] Good N et al, 'User Choices and Regret: Understanding Users' Decision Process about Consensually Acquired Spyware' (2006) 2(2) A Journal of Law & Policy for the Information Society 283, 323.
[110] Moores T, 'Do Consumers Understand the Role of Privacy Seals in E-Commerce?' (2005) 48(3) Communications of the ACM 86, 89-90.
[111] Böhme R and Köpsell S, 'Trained to Accept? A Field Experiment on Consent Dialogs' (2010) Proceedings of the SIGCHI Conference on Human Factors in Computing Systems 2403, 2406.
[112]Acquisti A, Brandimarte L, and Loewenstein G, 'Misplaced confidences: Privacy and the control paradox' (2012) Social Psychological and Personality Science, 1, p 6.

Because privacy choices are context-dependent, caution is needed when drawing conclusions about the effect of biases. One bias might influence a privacy decision in one direction, while another bias might influence the decision in another direction.[113] Still, it would be naïve to ignore behavioural sciences when making laws that rely, in part, on the decisions of people whose privacy the law aims to protect.

Biases can lead to behavioural market failures. These are 'market failures that complement the standard economic account and that stem from the human propensity to err.'[114] Apart from questions of fairness, such behavioural market failures decrease social welfare, in the same way as conventional market failures. 'Free markets', explains Sunstein, 'may well reward sellers who attempt to exploit human errors. In identifiable cases, those who do *not* exploit human errors will be seriously punished by market forces, simply because their competitors are profiting from doing so.'[115] Several authors conclude there is a behavioural market failure regarding online privacy. Firms would not stay in business if they did not exploit people's biases. As Strandburg puts it, '[t]he behavioral advertising business model seems almost designed to take advantage of (…) bounded rationality.'[116]

---

[113] Acquisti A and Grossklags J, 'What Can Behavioral Economics Teach Us About Privacy?' in Acquisti A et al (eds), *Digital Privacy: Theory, Technologies and Practices* (Auerbach Publications, Taylor and Francis Group 2007), p 371-374.

[114] Sunstein CR, *Why Nudge? The Politics of Libertarian Paternalism* (Yale University Press 2014), p 16.

[115] ibid, p 11.

[116] Strandburg KJ, 'Free Fall: the Online Market's Consumer Preference Disconnect' (2013) University of Chicago Legal Forum 95, p 149. See also Acquisti A, 'The economics of personal data and the economics of privacy (background paper conference: The Economics of Personal Data and Privacy: 30 Years after the OECD Privacy Guidelines)' (2010) <www.oecd.org/internet/ieconomy/46968784.pdf> accessed 4 February 2014, p 6.

# 5 How to Improve Privacy Protection?

Considering the limited potential of informed consent as a privacy protection measure, I argue that policymakers should use a combined approach of empowering and protecting the individual. Compared to the current approach, policymakers should focus more on protection.

This chapter distinguishes empowerment and protection rules to structure the discussion, but the distinction is not a formal legal distinction. The chapter uses rules that aim for *data subject control* and rules that aim for *empowerment* roughly interchangeably. Examples of empowerment rules are default rules that give the data subject the choice to allow data processing or not, such as informed consent requirements.[117] Other empowerment rules aim to make data processing transparent for the data subject, or grant data subjects rights, for instance to access and correct their data.[118]

Protection rules are generally mandatory. They always apply, irrespective of whether the data subject has consented to processing. For instance, under data protection law firms must always secure the data they process.[119] Another example of data protection law's aim to protect people is the existence of independent Data Protection Authorities that oversee compliance with the rules, as required by the Charter of Fundamental Rights of the European Union.[120]

## 5.1 Individual Empowerment

How could the law improve empowerment of the individual? To reduce the information asymmetry in the area of behavioural targeting, data protection law's transparency principle should be enforced more strictly. The Working Party says privacy policies and consent requests must be phrased in a clear and comprehensible manner.[121] The European Commission proposal for a Data Protection Regulation codifies this requirement.[122] Such a rule could discourage firms from using legalese in privacy policies, and would make it easier for DPAs to intervene when a firm uses a privacy policy or a consent request that is too vague. Given the fact that people currently would need several weeks per year to read privacy policies, such a rule would not be enough to ensure actual transparency. Still, the rule could help to lower the costs of reading privacy policies. And apart from data subjects, the press can also read privacy policies. A change in a firm's privacy policy could lead to media attention, and sometimes firms react to that.[123]

---

[117] See on the distinction between default and mandatory rules: Ayres I, 'Regulating Opt Out: An Economic Theory of Altering Rules' (2012).
[118] See article 10 and 11 (transparency) and article 12 (access and correction) of the Data Protection Directive.
[119] Article 17 (and 16) of the Data Protection Directive.
[120] Chapter 9 returns to the topic of rules that aim to protect the data subject.

In view of the limited effect that privacy policies have in informing people, interdisciplinary research is needed to develop tools to make data processing transparent in a meaningful way. Calo argues that we should not forget about transparency and informed consent, before better ways of presenting information have been tried.[124] Indeed, the current 'failure of mandated disclosure' does not prove that legal transparency requirements will always fail.[125] For instance, perhaps icons could be useful to communicate the data processing practices of firms. The European Commission encourages the use of icons,[126] and the European Parliament has proposed to require firms to use icons to inform people about data processing practices.[127]

---

[121] See Article 29 Working Party, 'Opinion 10/2004 on more harmonised information provisions' (WP 100), 25 November 2004.

[122] Art 11 of the European Commission proposal for a Data Protection Regulation (2012), see section 3.1.

[123] For instance, after attention in the press, Facebook offered people a way to opt out of their 'Beacon' service (Debatin B et al., 'Facebook and Online Privacy: Attitudes, Behaviors, and Unintended Consequences' (2009) 15(1) Journal of Computer-Mediated Communication 83).

[124] Calo MR, 'Against notice skepticism in privacy (and elsewhere)' (2011) 87(3) Notre Dame Law Review 1027.

[125] The phrase 'failure of mandated disclosure' is from: Ben-Shahar O and Schneider C, 'The Failure of Mandated Disclosure' (2011) 159 University of Pennsylvania Law Review 647, 650.

[126] European Commission, 'Communication from the Commission to the European Parliament and the Council on Promoting Data Protection by Privacy Enhancing Technologies (PETs), COM(2007)228 final, Brussels, 2 May 2007, par 4.3.2.

[127] See art 13(a), and the annex, of the proposal for a Data Protection Regulation, consolidated version after LIBE Committee vote, 22 October 2013, <www.janalbrecht.eu/fileadmin/material/Dokumente/DPR-Regulation-inofficial-consolidated-LIBE.pdf> accessed 30 June 2014. To me, the proposed six icons do not seem very clear. But it is possible that after a while, people would start to recognise the icons.

### *Consent for tracking technologies*

Human attention is scarce, and too many consent requests can overwhelm people. Therefore, the scope of article 5(3) of the e-Privacy Directive is too broad. Article 5(3) requires consent for storing or accessing information on a user's device. This means the provision also requires consent for some cookies that are not used to collect detailed information about individuals. But there is little reason to ask consent for innocuous practices. For instance, perhaps certain types of cookies that are used for website analytics could be exempted from the consent requirement, provided they do not threaten privacy and are not used to construct profiles of people.[128]

It would probably be better if the consent requirement for tracking were phrased in a more technology neutral way. The law could require consent for the collection and further processing of personal data, including pseudonymous data, for behavioural targeting and similar purposes – regardless of the technology that is used. Phrasing the rule in a more technology neutral way could also mitigate another problem. In some ways the scope of article 5(3) may be too narrow. For instance, it is unclear to what extent article 5(3) applies if firms use device fingerprinting for behavioural targeting.[129]

### *Privacy nudges*

Scholars have started to explore the possibilities for privacy nudges.[130] For example, Wang et al examine whether it is possible to help users of social network sites to avoid posting messages that they later regret.[131] A pop-up could warn people who post a status update on Facebook about how many people will be able to see that message. Such a warning can be made more forceful, for instance by including pictures of people who can see the post. Perhaps somebody would not post that 'drunk' picture if a pop-up reminds her that her boss can see the picture too. The researchers conclude 'that privacy nudges can potentially be a powerful mechanism to help some people avoid unintended disclosures.'[132] Balebako et al have explored 'nudging users towards privacy on mobile devices', to help people with decisions regarding the sharing of location data.[133] There is also research on nudging people to avoid installing privacy-invasive smart phone apps.[134]

---

[128] The Data Protection Directive has the balancing provision for such innocuous practices (art 7(f)). The Working Party argues for the introduction of an exception in art 5(3) for certain types of innocuous analytics cookies (Article 29 Working Party, 'Opinion 04/2012 on Cookie Consent Exemption' (WP 194) 7 June 2012).

[129] See on device fingerprinting: Acar G et al, 'The Web never forgets: Persistent tracking mechanisms in the wild' (Draft, 10 August 2014) <https://securehomes.esat.kuleuven.be/~gacar/persistent/the_web_never_forgets.pdf> accessed 14 August 2014.

[130] Acquisti A, 'Nudging Privacy. The behavioral economics of personal information' (2009) 7(6) IEEE Security & Privacy 82.

[131] Wang Y and others, 'The Second Wave of Global Privacy Protection: From Facebook Regrets to Facebook Privacy Nudges' (2013) 74 Ohio State Law Journal 1307, p 1307.

Setting defaults is a classic example of nudging. The status quo bias suggests that requiring opt-in consent for tracking could nudge people towards disclosing fewer data.[135] Hence, if the goal is protecting privacy, behavioural sciences insights suggest that the law should require opt-in systems for valid consent. This implies that the existing rules regarding consent should be enforced. As noted, the European Commission proposal for a Data Protection Regulation tightens the requirements for consent.[136] The proposal also codifies the Working Party's view that a consent request may not be hidden in a privacy policy or in terms and conditions.[137]

***Tracking walls***

There is a problem when policymakers prescribe default settings to make behavioural targeting firms nudge people towards disclosing less data. Such firms may have an incentive to collect as much information as possible. As Willis notes, firms have many ways to persuade people to opt in to tracking.[138] It is hard for policymakers to make firms use behaviourally informed instruments, if firms do not want to nudge people in the same direction as the policymaker. Sunstein puts it as follows: 'if regulated institutions are strongly opposed to a default rule and have easy access to their customers, they may well be able to use a variety of strategies, including behavioral ones, to encourage people to move in the direction the institutions prefer.'[139] For instance, firms can offer take-it-or-leave-it choices, such as tracking walls on websites. Hence, even if firms offered transparency and asked prior consent for behavioural targeting, people might still feel they must consent.

---

[132] Idem, p 1334.
[133] Balebako R and others, 'Nudging users towards privacy on mobile devices' CHI 2011 workshop article <www.andrew.cmu.edu/user/pgl/paper6.pdf> accessed 18 August 2014.
[134] Choe EK et al, 'Nudging people away from privacy-invasive mobile apps through visual framing' in *Human-Computer Interaction–INTERACT 2013* (Springer 2013).
[135] See: Sunstein CR, 'Deciding By Default' (2013) 162(1) University of Pennsylvania Law Review 1.
[136] Art 4(8) of the European Commission proposal for a Data Protection Regulation (2012). See section 3.1 of this chapter.
[137] Article 29 Working Party, 'Opinion 15/2011 on the definition of consent' (WP 187) 13 July 2011, p 33-35
[138] Willis LE, 'Why Not Privacy by Default?' (2013) 29 Berkeley Technology Law Journal 61.

Should the law do anything about take-it-or-leave-it choices regarding the enjoyment of privacy when using websites and other Internet services? Some have suggested that tracking walls and similar take-it-or-leave-it-choices should be prohibited.[140] Another option would be to ban such take-it-or-leave-it choices in certain contexts. (The next section returns to the topic of context-specific rules.)

It has also been suggested that the law should require firms to offer a tracking-free version of their services, which has to be paid for with money.[141] Such a rule would enable people to compare the prices of websites. Now the 'price' of a website is usually hidden because people do not know which information about them is captured, nor how it will be used.[142] A requirement for firms to offer a tracking-free but paid-for version of their service would be less protective of privacy than a ban on tracking walls. Myopia might lead most people to choose for the so-called free version, because they focus on the short-term loss of paying for a service, also if this means they have to consent to behavioural targeting, contrary to earlier plans.[143] Furthermore, many say it is 'extortion' if they have to pay for privacy.[144]

---

[139] Sunstein CR, *Simpler: The Future of Government* (Simon and Schuster 2013), p 119.

[140] See e.g.: Irion K and Luchetta G, 'Online Personal Data Processing and EU Data Protection Reform' (CEPS Task Force Report of the CEPS Digital Forum 2013) <www.ivir.nl/publications/irion/TFR_Data_Protection.pdf> accessed 19 August 2014, p 78.

[141] Traung P, 'The Proposed New EU General Data Protection Regulation: Further Opportunities' (2012)(2) Computer Law Review international 33, p 42; Irion K and Luchetta G, 'Online Personal Data Processing and EU Data Protection Reform' (CEPS Task Force Report of the CEPS Digital Forum 2013), p 38.

[142] Helberger N, 'Freedom of Expression and the Dutch Cookie-Wall' (2013), Institute for Information Law <www.ivir.nl/publications/helberger/Paper_Freedom_of_expression.pdf> accessed 5 November 2013, p 19.

[143] See on the attraction of 'free' offers: Ariely D, *Predictably Irrational* (Harper 2008), ch 3; Hoofnagle CJ and Whittington JM. 'The price of 'free': accounting for the cost of the Internet's most

***Why still aim for empowerment?***

The behavioural sciences analysis suggests that the practical problems with informed consent are immense. So why still aim for empowerment? First, it seems unlikely that Europe would ever adopt a data protection regime without any role for informed consent, if only because the European Union Charter of Fundamental Rights lists consent as one of the possible legal grounds for personal data processing.[145] In addition, people's tastes differ. Some people would approve of a certain behavioural targeting practice, while others would not. Regulation with an informed consent provision has the advantage of respecting people's individual preferences. Taking away *all* privacy choices from the individual would probably make the law unduly paternalistic. Furthermore, it does not seem feasible to define all beneficial or all harmful data processing activities in advance. Indeed, several scholars that are extremely sceptical of informed consent as a privacy protection measure still conclude that a legal privacy regime without any role for informed consent is neither feasible nor desirable.[146]

Relying on informed consent, in combination with data protection law's other safeguards, will probably remain the appropriate approach in many circumstances. For those cases, transparency and consent should be taken seriously. Fostering individual control over personal information alone will not suffice to protect privacy in the area of behavioural targeting. But some improvement must be possible, compared to the current situation of almost complete lack of individual control over personal information.

---

popular price (forthcoming (2014) UCLA Law Review 61(3))' <http://ssrn.com/abstract=2235962> accessed 30 June 2014.

[144] Cranor LF and McDonald AM. 'Beliefs and Behaviors: Internet Users' Understanding of Behavioral Advertising (38th Research Conference on Communication, Information and Internet Policy (Telecommunications Policy Research Conference)) (2 October 2010) <http://ssrn.com/abstract=1989092> accessed 30 June 2014, p 27.

[145] See art 8(2) of the European Union Charter of Fundamental Rights.

[146] See e.g. Barocas S and Nissenbaum H. 'On Notice: the Trouble with Notice and Consent (Proceedings of the Engaging Data Forum: The First International Forum on the Application and Management of Personal Electronic Information)' (October 2009) <www.nyu.edu/pages/projects/nissenbaum/papers/ED_SII_On_Notice.pdf> accessed 30 June 2014; Solove DJ, 'Privacy Self-Management and the Consent Dilemma' (2013) 126 Harvard Law Review 1879.

In conclusion, regarding the requirements for valid consent, the formal legal framework is essentially in line with behavioural sciences insights. Firms are not allowed to infer consent from mere silence – and should not be allowed to do so. But even if firms offered transparency and asked for opt-in consent for tracking, the issue of take-it-or-leave-it choices and tracking walls would remain. As long as the law allows take-it-or-leave-it choices, opt-in systems will not be effective privacy nudges.

### 5.2 Individual Protection

A second legal approach to improve privacy protection in the area of behavioural targeting involves *protecting,* rather than empowering, people. If fully complied with, the data protection principles could give reasonable privacy protection in the area of behavioural targeting, even if people agreed to consent requests.

Of course, the Data Protection Directive is only relevant if the practice of behavioural targeting is found to come within the directive's scope. This will be the case if pseudonymous data, such as the data that are used for behavioural targeting, are seen as personal data. Hence, from a normative perspective, data protection law should apply to behavioural targeting, including when firms use pseudonymous data. Apart from that, a sensible interpretation of data protection law implies that data that are used to 'single out' a person should be seen as personal data.[147]

While consent plays an important role in data protection law, its role is also limited. Consent can provide a legal ground for personal data processing. However, if a firm has a legal ground for processing, the other data protection provisions still apply.[148] Those provisions are mandatory. The data subject cannot waive the safeguards or deviate from the rules by contractual agreement. For example, even after a firm obtains an individual's consent, data protection law does not allow excessive personal data processing. And it follows from the purpose limitation principle that personal data must be collected for specified purposes, and may not be used for unexpected purposes.[149] Hence, a contract between a firm and a data subject would not be enforceable if it stipulated that the firm does not have to secure the personal data, or can use the data for new purposes at will.

---

[147] See section 3.1 of this chapter, and in detail on the material scope of data protection law: chapter 5 of my forthcoming PhD thesis (see note 1 of this chapter).

[148] See: CJEU, Case C-131/12, Google Spain SL and Google Inc. v. Agencia Española de Protección de Datos and Mario Costeja González, not yet published, par 71: if there is a legal ground for processing (art 7), the other provisions of the Data Protection Directive still apply.

The data minimisation principle, if effectively enforced, is an example of a data protection principle that could protect privacy, also after people consent to behavioural targeting. The Data Protection Directive says data processing must be 'not excessive' in relation to the processing purpose, and it follows from the Directive's structure that this requirement also applies if the processing is based on the data subject's consent.[150] The vast scale of data processing for behavioural targeting aggravates the chilling effects, and the lack of individual control over personal information. And large-scale data storage brings risks, such as data breaches. Compliance with the data minimisation principle could mitigate such privacy problems. Policymakers should explicitly codify that the data subject's consent does not legitimise disproportionate data processing. Such a rule could remind firms that consent does not give them *carte blanche* to collect personal information at will, and that a DPA could intervene if they did.

Perhaps the law could prohibit storing data for behavioural targeting longer than a set period of, to give an example, two days. Such a hard and fast rule would provide more legal certainty than the general data minimisation principle. Compared to estimating when the data minimisation principle requires deletion, complying with a maximum retention period of two days is easy for firms. As an aside, it is unclear whether storing tracking data for longer than a few days helps much to improve the click-through rate on ads.[151]

Data protection law's transparency principle can be interpreted as a prohibition of surreptitious data processing. Hence, while the last section discussed the transparency principle as a means to empower the individual, the principle could also be seen as more prohibitive. With some behavioural targeting practices, it would be hard for a website publisher to comply with data protection law's transparency requirements, even if it tried its best. For example, some ad networks allow other ad networks to buy access to individuals (identified through cookies or other identifiers) by bidding on an automated auction.[152] In such situations, the website publisher does not know in advance which ad networks will display ads on its site, and which ad networks will track its website visitors. In data protection parlance: the publisher does not know who the joint data controllers are.[153] Neither does the publisher know for which purposes the ad networks will use the data they collect. As noted, the Directive obliges data controllers to provide a data subject information about their identity, the processing purpose, and all other information that is necessary to guarantee fair processing.[154] Therefore, it is hard to see how the publisher could comply with the law's transparency requirements. If a publisher cannot give data subjects the information that is required by the Data Protection Directive, the processing is not allowed – and should not be allowed. Policymakers should make more explicit that processing is prohibited, unless firms can comply with the transparency principle.

---

[149] Art 17 and art 6(1)(b) of the Data Protection Directive.

[150] Art 6(1)(c) and art 6(1)(e) of the Data Protection Directive. In data protection literature, authors tend to speak of data minimisation. One could also speak of proportionality.

[151] See Strandburg KJ, 'Free Fall: the Online Market's Consumer Preference Disconnect' (2013) University of Chicago Legal Forum 95, p 104-105.

Data protection law has a stricter regime for 'special categories of data', such as data revealing race, political opinions, health, or sex life. The use of special categories of data for behavioural targeting and other types of direct marketing is prohibited, or, depending on the national implementation law, only allowed after the data subject's 'explicit' consent.[155] Strictly enforcing the existing rules on special categories of data could reduce privacy problems such as chilling effects. Let us take health data as an example. People might be hesitant to look for medical information on the web if they fear leaking information about their medical condition. The rules on special categories of data could be interpreted in such a way that the collection context is taken into account. For instance, arguably tracking people's visits to websites with medical information should be seen as the processing of special categories of data, as firms could infer data regarding health from such tracking information.[156] Furthermore, policymakers should consider banning the use of any health related data for behavioural targeting. The privacy risks involved in using health data for behavioural targeting seem to outweigh the possible societal benefits from such practices.

---

[152] Castelluccia C, Olejnik L and Minh-Dung T, 'Selling Off Privacy at Auction' (2013) Inria <www.inrialpes.fr/planete/people/lukasz/rtbdesc.html> accessed 30 June 2014.

[153] The Working Party says ad networks and website publishers are often joint data controllers, as they jointly determine the purposes and means of the processing. See: Article 29 Working Party, 'Opinion 2/2010 on online behavioural advertising' (WP 171), 22 June 2010, p 11

[154] Art 10 and 11 of the Data Protection Directive (see section 4.2 of this chapter). See also: article 5(3) of the e-Privacy Directive.

[155] Art 8 of the Data Protection Directive.

The law could also prohibit take-it-or-leave-it choices in some circumstances or contexts.[157] For instance, public service broadcasters often receive public funding, and they have a special role in informing people. But if people fear their behaviour is being monitored, they might forego the use of public service media.[158] To reduce such chilling effects, policymakers should prohibit public service broadcasters to use tracking walls or similar take-it-or-leave-it choices. Policymakers could also go one step further, and prohibit all third party tracking for behavioural targeting on public service media. More generally it is questionable whether it is appropriate for public sector websites to allow third party tracking for behavioural targeting – even when people consent. It is not evident why the public sector should facilitate tracking people's behaviour for commercial purposes. Therefore, policymakers should consider prohibiting all tracking for behavioural targeting on public sector websites.

***Using transaction costs strategically***

Policymakers could also use an intermediate option between default rules that aim to empower people and mandatory protective rules. Policymakers could use transaction costs strategically.[159] As noted, marketers understand the importance of transaction costs – and sometimes use them strategically. Opting out of behavioural targeting often takes more effort than opting in. On the website Youronlinechoices, managed by the Interactive Advertising Bureau, it takes three clicks and a waiting period to opt out of receiving behaviourally targeted ads.[160] In principle, policymakers could do something similar.

---

[156] See ECJ, Case C-101/01 Lindqvist, [2003] ECR I-12971, para 50: 'the expression "data concerning health" (…) must be given a wide interpretation'.

[157] See on the importance of context for privacy: Nissenbaum H, *Privacy in context: technology, policy, and the integrity of social life* (Stanford Law Books 2010).

[158] See: Helberger N, 'Freedom of Expression and the Dutch Cookie-Wall' (2013), Institute for Information Law <www.ivir.nl/publications/helberger/Paper_Freedom_of_expression.pdf> accessed 18 August 2014.

[159] Thanks for Oren Bar-Gill for introducing me to this idea.

For example, policymakers could strengthen a nudge that consists of setting a default by adding transaction costs, thereby making the default stickier.[161] Perhaps one mouse click could be required to give consent to relatively innocuous kinds of tracking. Three mouse clicks could be required for more worrying practices. 'Sticky defaults', says Ayres, 'should be thought of as an intermediate category falling between ordinary defaults and traditional mandatory rules.'[162] Transaction costs could come in different shades, to introduce different degrees of stickiness for the default. In theory the law could require a thirty second waiting period, a phone call, or a letter by registered mail to opt in to certain practices.[163]

But caution is needed if policymakers consider adding friction to consent procedures in the area of behavioural targeting. A legal regime that adds transaction costs and allows firms to offer take-it-or-leave-it choices could lead to an unpleasant situation. Website publishers could use tracking walls, also if policymakers required three mouse clicks for consent. People would not enjoy clicking three times 'I agree' if they want to visit a website, and accept they have to agree to tracking. With that caveat, the conclusion still stands: the distinction between mandatory rules and opt-in systems (default rules) is not a black and white issue. In principle policymakers have a range of options.

In conclusion, enforcing and tightening the data protection principles could help to protect privacy in the area of behavioural targeting. An important topic that falls outside this chapter's scope is how enforcement of European data protection law could be improved, in particular when firms are based outside Europe.[164] Anyhow, enforcing data protection law may not be enough to protect privacy in the area of behavioural targeting. If society is better off when certain behavioural targeting practices do not happen, policymakers should consider banning them. Agreeing on prohibitions would be hard. But that should not be a reason to ignore this legal tool.

---

[160] In a non-scientific test, I had to wait forty-five seconds. First I had to choose a country (click 1), then I had to click on "your ad choices" (click 2). Next I had to wait until the website contacted the participating ad networks. Then I could opt out of receiving targeted advertising (click 3). For several ad networks the website gave an error message (Youronlinechoices <http://www.youronlinechoices.com/ma/faqs#21> accessed 30 June 2014).

[161] If a nudge is made stronger by using transaction costs strategically, it might not count as a 'nudge' anymore, since it is not 'easy and cheap to avoid' (Sunstein CR and Thaler RH, Nudge: Improving Decisions about Health, Wealth, and Happiness (Yale University Press 2008), p 6). I will leave this complication aside.

[162] Ayres I, 'Regulating Opt Out: An Economic Theory of Altering Rules' (2012), p 2087.

[163] See for a discussion: Willis LE, 'Why Not Privacy by Default?' (2013) 29 Berkeley Technology Law Journal 61, p 121-128.

## 6 Conclusion

To protect privacy in the area of behavioural targeting the European Union mainly relies on the consent requirement for the use of tracking technologies in the e-Privacy Directive, and on general data protection law. With informed consent requirements, the law aims to empower people to make choices in their best interests. But behavioural studies cast doubt on the effectiveness of such an approach.

There is no silver bullet to improve privacy protection in this area. While current regulation emphasises empowerment and informed consent, without much reflection on practical issues, I argue for a combined approach of protecting and empowering people. To improve individual empowerment, the data protection rules should be tightened, and should be enforced more strictly. For example, long unreadable privacy policies should not be accepted. Aiming for empowerment will not suffice to protect privacy. Nevertheless, some improvement must be possible, as today personal data are generally captured and used without meaningful transparency or consent.

Policymakers could also try to nudge Internet users towards disclosing less data. For instance, policymakers could require firms to obtain the individual's opt-in consent for tracking. The discussion about privacy defaults has been going on for the past 25 years in Europe. Marketers have argued that they should be allowed to use opt-out systems to obtain 'implied' consent for direct marketing, and, more recently, for tracking cookies. In line with legal doctrine, European Data Protection Authorities say consent requires an expression of will, which generally calls for opt-in procedures. The debate essentially concerns the direction of a nudge: who benefits from the status quo bias, the firm or the data subject? However, such privacy nudges run into problems. As long as the law allows firms to offer-take-it-or-leave-it choices, firms can easily persuade people to agree to tracking.

---

[164] See: European Agency for Fundamental Rights, 'Data Protection in the European Union: the Role of National Data Protection Authorities' (2010) <http://fra.europa.eu/sites/default/files/fra_uploads/815-Data-protection_en.pdf> accessed 16 October 2013.

I argue that policymakers should focus more on protecting people. After all, the European Court of Human Rights requires privacy protection that is 'practical and effective, not theoretical and illusory.'[165] While the role of informed consent in data protection law is important, that role is limited at the same time. People cannot waive data the safeguards of protection law, or contract around the rules. The protective data protection principles should be enforced more strictly. For example, even after consent, excessive data processing is not allowed – and should not be allowed. But enforcing data protection law will not be enough to protect privacy. In addition to data protection law, more specific rules regarding behavioural targeting are needed. And if society is better off if certain behavioural targeting practices don't happen, policymakers should consider banning them.

* * *

[165] ECtHR Christine Goodwin v UK (App 28957/95) (2002) 35 EHRR 18, par 74.